\documentclass[noshowpacs,amsmath,
twocolumn,
superscriptaddress,
8pt,
aps,prb]{revtex4-2}
\usepackage{setspace}
\usepackage{amsmath}
\usepackage{multirow}
\usepackage{graphicx}
\usepackage{booktabs}
\usepackage{upgreek}

\renewcommand{\figurename}{\textbf{Fig.}}
\usepackage{verbatim}
\usepackage{amsfonts}
\usepackage{amssymb}
\usepackage{epstopdf}
\usepackage{dcolumn}
\usepackage{xcolor}
\usepackage{verbatim}
\usepackage{siunitx}
\usepackage{makecell}
\usepackage{float}
\usepackage{lineno}

\setcitestyle{super}
\DeclareGraphicsExtensions{.pdf,.eps,.png,.jpg,.mps}

\begin{document}

\title{Integrated optomechanical ultrasonic sensors with nano-Pascal-level sensitivity}




\author{Xuening Cao$^{1,2,*}$, Hao Yang$^{1,2,*}$, Min Wang$^{1,2,*}$, Zhi-Gang Hu$^{1,2}$, Zu-Lei Wu$^{1,2}$, Yuanlei Wang$^{3,1}$, Jian-Fei Liu$^{1,2}$, Xin Zhou$^{1,2}$, Jincheng Li$^{1,5}$, Chenghao Lao$^{3}$, Qi-Fan Yang$^{3,4}$, and Bei-Bei Li$^{1,2,6,\dagger}$\\
$^1$Beijing National Laboratory for Condensed Matter Physics, Institute of Physics, Chinese Academy of Sciences, Beijing 100190, China\\
$^2$University of Chinese Academy of Sciences, Beijing 100049, China\\
$^3$State Key Laboratory for Artificial Microstructure and Mesoscopic Physics and Frontiers Science Center for Nano-optoelectronics, School of Physics, Peking University, Beijing 100871, China\\
$^4$Collaborative Innovation Center of Extreme Optics, Shanxi University, Taiyuan 030006, Shanxi, China\\
$^5$School of Space and Earth Sciences, Beihang University, Beijing 100191, China\\
$^6$Songshan Lake Materials Laboratory, Dongguan 523808, Guangdong, China\\
$^{*}$These authors contributed equally to this work.\\
$^{\dagger}$Corresponding author: libeibei@iphy.ac.cn}

\date{}

\maketitle
{\bf\noindent Ultrasonic sensors are widely used for object detection and localization in underwater and biological settings. The operational range and spatial resolution are inherently limited by sensor sensitivity, in which conventional piezoelectric transducers have been overwhelmed by advanced photonic sensors. Here, we demonstrate an optomechanical ultrasonic sensor integrated into a photonic platform, which comprises a suspended SiO$_2$ membrane embedded with a high-$Q$ Si$_3$N$_4$ microring resonator. By exploiting simultaneous optical and mechanical resonances, the sensor achieves a record low noise-equivalent pressure (NEP) of 218~nPa/$\sqrt{\rm{Hz}}$ at 289~kHz in air and 9.6~nPa/$\sqrt{\rm{Hz}}$ at 52~kHz in water. We demonstrate its versatility through photoacoustic gas spectroscopy in air and underwater ultrasound imaging, achieving a minimum detectable C$_2$H$_2$ concentration of 2.9~ppm (integration time 1~s) and an imaging resolution of 1.89~mm, respectively. Our work represents a significant advancement in compact CMOS-compatible ultrasound sensing, unlocking new possibilities in biomedical imaging, environmental monitoring, industrial testing, and underwater communications.}

\begin{figure*}[ht!]
\centering
\includegraphics{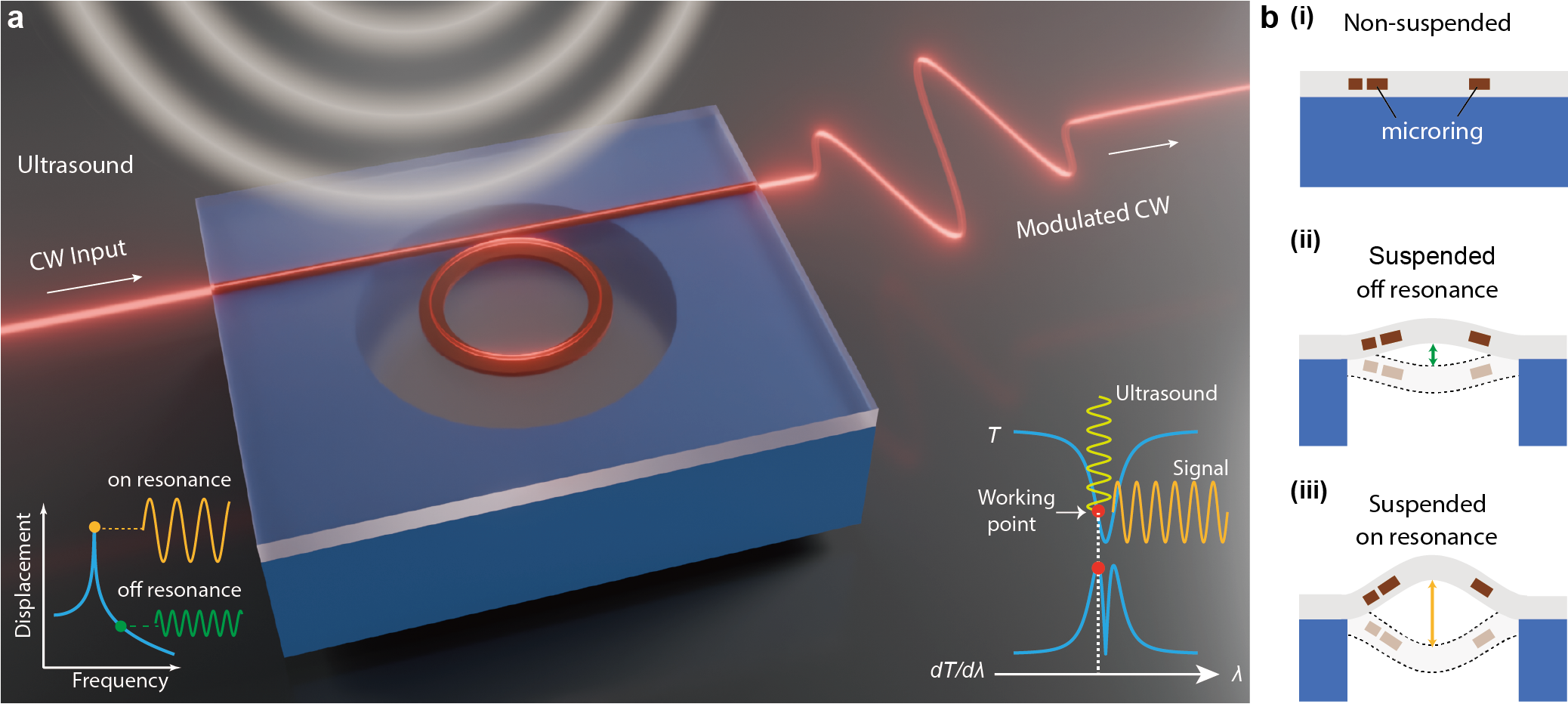}
\caption{\textbf{Schematic and working principle of the integrated ultrahigh-sensitivity optomechanical ultrasonic sensor.}
\textbf{a} Schematic of the ultrasonic sensor, where the intensity of the continuous wave (CW) input laser is modulated by the ultrasound. Left inset: Schematic of the mechanical resonance-enhanced displacement response. Right inset: Optical readout principle of the sensor, illustrating the optical resonance-enhanced sensitivity.
\textbf{b} Schematic comparison of integrated microring resonator-based ultrasonic sensors with three configurations: (i) non-suspended structure, (ii) suspended structure (off mechanical resonance), and (iii) suspended structure (on mechanical resonance), illustrating the enhanced response of the suspended structure with mechanical resonance.}
\label{Fig1}
\end{figure*}

Ultrasound sensing has become a cornerstone of modern diagnostic and monitoring technologies, with crucial applications spanning medical imaging \cite{wearableUS1, wearableUS2}, non-destructive testing \cite{24,25}, underwater acoustics \cite{27}, and industrial process monitoring \cite{23,26}. Conventional piezoelectric transducers \cite{E1, E2-3, E4-5}, though widely used, face critical challenges in miniaturization, integration, and sensitivity, limiting their applications in scenarios that require compact designs or high resolutions \cite{10,12}. 
These constraints are particularly evident in miniaturized systems such as portable medical devices \cite{18} and intravascular ultrasound probes \cite{17, FBG11}, where sensor arrays are often restricted to sub-centimeter scale to meet spatial and operational requirements. Recent advances in micromachined sensors, such as piezoelectric micromachined ultrasound transducers (PMUTs) \cite{E6} and capacitive micromachined ultrasound transducers (CMUTs) \cite{CMUT1, CMUT2, CMUT3-5}, have improved array density and electronic integration. However, these devices remain susceptible to electromagnetic interference and exhibit inadequate sensitivities required for next-generation applications. For instance, their inability to operate close to high-voltage transmission lines hinders the monitoring of corona discharge noise \cite{19}. Furthermore, their insufficient sensitivity poses significant challenges in applications like cranial ultrasound imaging \cite{20,21}.

Optical microcavity-based ultrasonic sensors have emerged as a transformative alternative \cite{Cao2024, E4-5, 13}, offering unprecedented sensitivity, immunity to electromagnetic interference, broad bandwidth, and chip-scale integration capability. Fiber-based microcavity sensors, including Fabry-Pérot cavities \cite{FP3, FP6, FP13}, Bragg gratings \cite{FBG5, FBG11, FBG4, FBG7}, and microsphere cavities \cite{WGM18_sphere, WGM26_sphere, WGM27_sphere}, have shown particular promise in endoscopy applications but face scalability challenges due to limited arraying capabilities and reliance on bulky optical components \cite{FP8}. 
The growing demand for miniaturized, high-performance sensing systems has spurred significant research toward chip-scale sensing platforms \cite{1}, where CMOS-compatible microring resonators show significant promise. Various microring resonator platforms, fabricated from polymer \cite{WGM6_ring, WGM9_ring, WGM14_ring}, silicon \cite{WGM11_ring, WGM24_ring}, and chalcogenide \cite{WGM31_ring}, have been developed for ultrasound sensing. Despite their broad detection bandwidth, these integrated sensors achieved noise equivalent pressures (NEPs) only in the mPa/$\sqrt{\rm{Hz}}$ range, primarily due to their modest optical quality ($Q$) factors and restricted mechanical displacements caused by the thick substrates.

Optomechanical sensors offer a powerful approach for ultrasensitive ultrasound detection by harnessing the strong interaction between light and mechanical vibrations in high-$Q$ microresonators. Suspended microdisk resonators have achieved unprecedented low NEPs at the \textmu Pa/$\sqrt{\rm{Hz}}$ level \cite{WGM15_disk, WGM29_disk, WGM32_disk}. However, the reliance on fiber-taper coupling presents significant challenges in practical applications, including alignment complexity, reduced robustness, and difficulty working in aqueous environments. Here, we develop a novel integrated, mass-produced, ultrasensitive optomechanical ultrasonic sensor with nano-Pascal-level sensitivity and further demonstrate the sensor's versatility through photoacoustic gas spectroscopy and underwater ultrasound imaging. Beyond prototype demonstrations, this fully integrated, CMOS-compatible, highly sensitive ultrasonic sensor enables scalable production and paves the way for future multi-scene sensing applications, advancing medical diagnostics and industrial precision measurements.

\begin{figure*}
    \centering
    \includegraphics{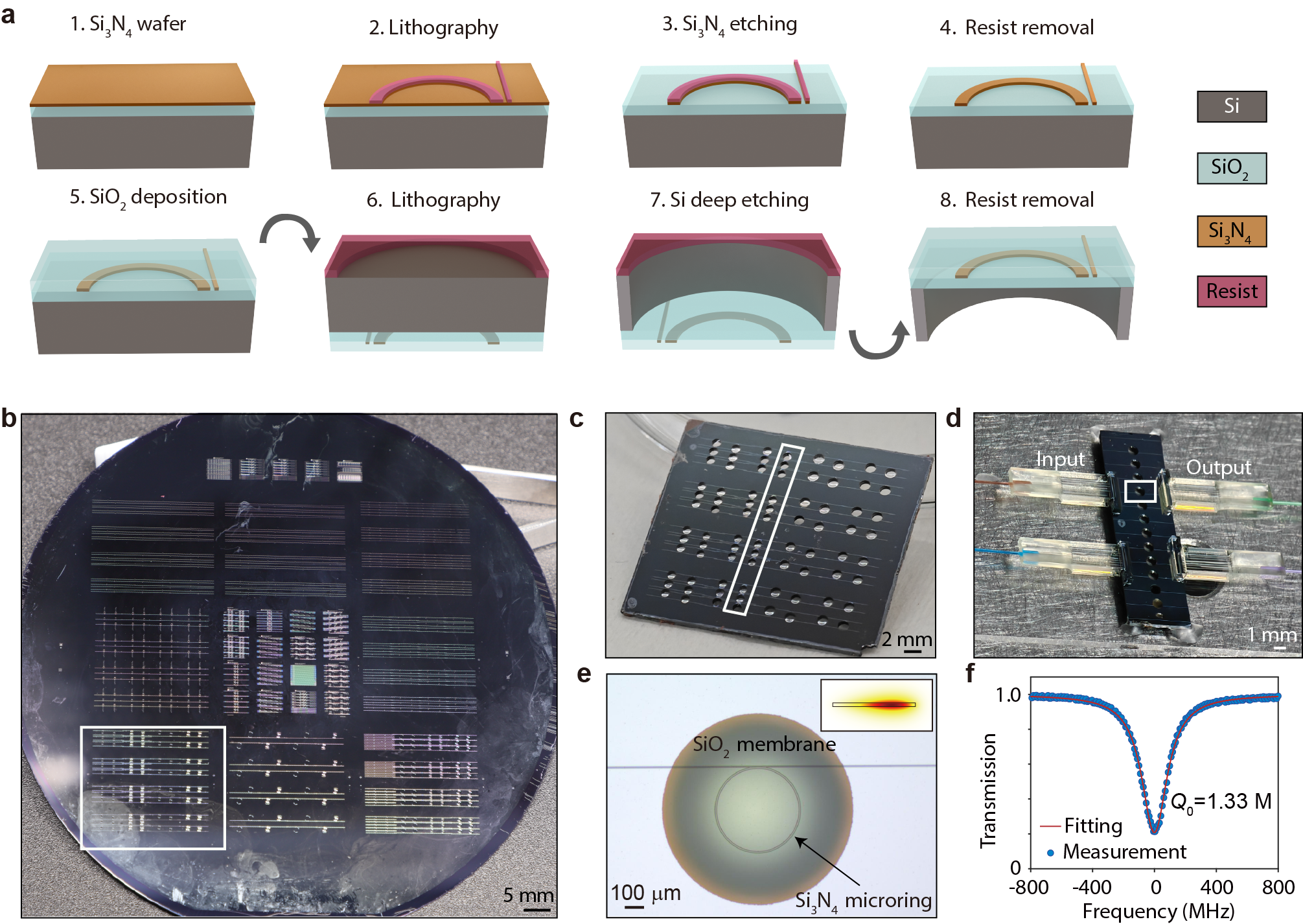}
    \caption{\textbf{Sensor fabrication.} 
    \textbf{a} Fabrication process of the sensor. \textbf{b} Photograph of the processed 4-inch wafer. \textbf{c} Photograph of a 3~cm$\times$3~cm chip after Si deep etching. \textbf{d} photograph of the packaged sensor. \textbf{e} Optical microscopy image of the sensor, showing a SiO$_2$ membrane with an embedded Si$_3$N$_4$ microring resonator. Inset: Optical mode profile $|\textbf{E}|$, with $\textbf{E}$ denoting the electric field. \textbf{f} Normalized transmission spectrum of the microring resonator measured in water. }
    \label{Fig2}
\end{figure*}

\medskip
{\bf\noindent Results} 

\medskip
\begin{footnotesize}
\noindent{\bf Principle}
\end{footnotesize}

\noindent As depicted in Fig. \ref{Fig1}a, the proposed integrated ultrasonic sensor comprises a suspended SiO$_2$ membrane and a high-$Q$ Si$_3$N$_4$ microring resonator. The SiO$_2$ membrane is clamped at its periphery to the silicon substrate and suspended at its center to maximize mechanical displacement under ultrasound excitation. The Si$_3$N$_4$ microring is embedded within the SiO$_2$ membrane, enabling optomechanical readout by transducing the mechanical displacement of the membrane into optical resonance shifts of the ring resonator. To maximize the readout signal, we systematically optimize the geometries of the sensor, yielding final radii of 450~\textmu m for the membrane and 235~\textmu m for the microring resonator (see Methods and Extended Data Fig. \ref{FigED1}). As shown in the right inset of Fig. \ref{Fig1}a, by locking the frequency of an input continuous wave (CW) laser on the blue-detuned side of the optical resonance, the ultrasound-induced displacement is converted into the laser intensity modulation. The higher the optical $Q$ factor, the steeper the slope of the transmission spectrum, and the greater the readout sensitivity.

Mechanical resonance further amplifies the displacement of the SiO$_2$ membrane, significantly enhancing the sensitivity to acoustic signals. The bottom left inset of Fig. \ref{Fig1}a compares the displacement response of a mechanical oscillator at different frequencies, illustrating the amplification of the mechanical resonance. Figure \ref{Fig1}b presents a schematic comparison of integrated ultrasonic sensors with three configurations: (i) conventional non-suspended structure, suspended structure operating at (ii) off-mechanical and (iii) on-mechanical resonance frequencies, highlighting the mechanical resonance-enhanced sensor response. 

\begin{footnotesize}
\noindent{\bf Sensor design and fabrication}
\end{footnotesize}

\noindent We fabricate the devices at the wafer scale, with the process flow outlined in Fig. \ref{Fig2}a. The fabrication begins with electron beam lithography (EBL) and reactive ion etching (RIE) to create the Si$_3$N$_4$ microring resonator, with both processes optimized to achieve near-vertical sidewalls (Extended Data Fig. \ref{FigED2}). A SiO$_2$ cladding layer is deposited on the Si$_3$N$_4$ layer to preserve resonator optical performance, with Fig. \ref{Fig2}b showing a photograph of a full 4-inch wafer after the deposition. We then flip the wafer upside down and perform photolithography and deep reactive ion etching (DRIE) processes to selectively etch the silicon underneath Si$_3$N$_4$ microring resonators. Figure \ref{Fig2}c presents a photograph of a 3~cm$\times$3~cm chip after releasing the suspended SiO$_2$ membranes. For full integration and enhanced robustness, two mode conversion fibers are used to couple light into the microring resonator and are co-packaged with the device (Fig. \ref{Fig2}d). This results in a compact and portable ultrasonic sensor with reliable operation in diverse environments.

\begin{figure*}
    \centering
    \includegraphics{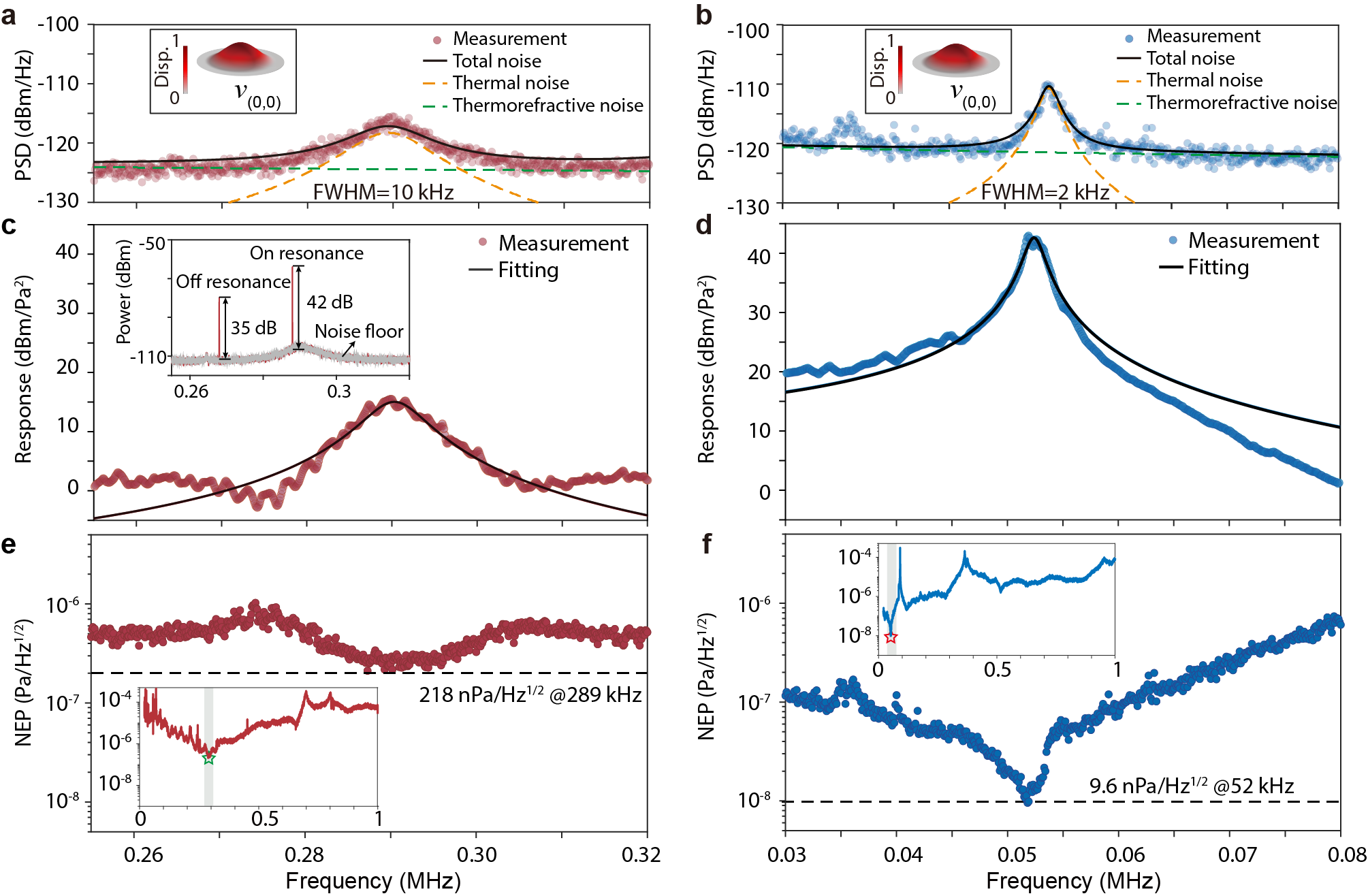}
    \caption{\textbf{Characterization of the ultrasonic sensor.} 
    \textbf{a,b} Noise power spectral densities (PSDs) of the $\nu_{(0,0)}$ mode in air (a) and water (b), with insets showing the $\nu_{(0,0)}$ mode profile of the SiO$_2$ membrane. FWHM: full width at half maximum.
    \textbf{c,d} Ultrasound response of the sensor at different frequencies in air (c) and water (d). Inset in (c): PSD of the sensor in air when the applied ultrasound frequency is on resonance (288~kHz) and off resonance (268~kHz) with the $\nu(0,0)$ mechanical mode, with the grey curve representing the noise floor. 
    \textbf{e,f} Noise-equivalent pressure (NEP) spectral densities around the $\nu_{(0,0)}$ mode in air (e) and water (f). The achieved minimum NEPs are 218~nPa/$\sqrt{\rm{Hz}}$ at 289~kHz in air and 9.6~nPa/$\sqrt{\rm{Hz}}$ at 52~kHz in water, respectively. Insets: NEP spectral densities between 20~kHz-1~MHz in air (e) and water (f).}
    \label{Fig3}
\end{figure*}

\begin{figure*}
\centering\includegraphics{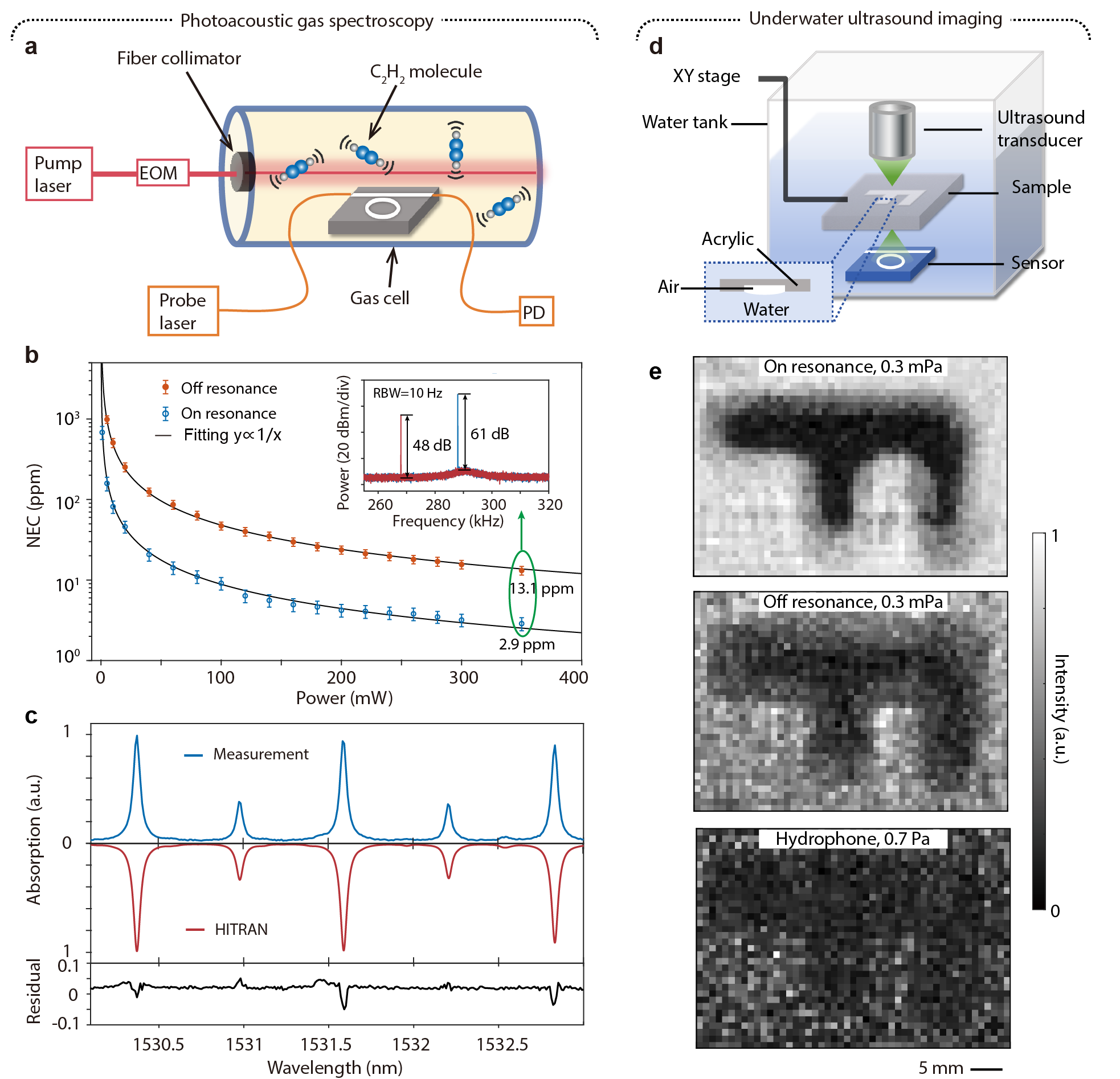}
\caption{\textbf{Demonstrations of photoacoustic gas spectroscopy and underwater ultrasound imaging.}
\textbf{a} Experimental setup for photoacoustic detection of gaseous acetylene (C$_2$H$_2$). EOM: electro-optic modulator, PD: photodetector. \textbf{b} Noise-equivalent concentration (NEC) as a function of pump power. Inset: Photoacoustic signal for 1\% C$_2$H$_2$ mixed with 99\% N$_2$ at 350~mW pump power. \textbf{c} Measured (blue) and HITRAN-simulated (red) absorption spectra of C$_2$H$_2$ from 1530~nm to 1533~nm. Bottom panel: Residual between experimental and simulated spectra. \textbf{d} Schematic of the underwater ultrasound transmission imaging system. Inset: Cross-sectional view of the sample structure. \textbf{e} (i,ii) Ultrasound images at 0.3~mPa pressure with an ultrasound frequency of (i) 517~kHz (on mechanical resonance) and (ii) 550~kHz (off mechanical resonance). (iii) Ultrasound image from a commercial hydrophone, with an ultrasound frequency of 517~kHz and pressure of 0.7~Pa.}
\centering
\label{Fig4}
\end{figure*}

\begin{footnotesize}
\noindent{\bf Device characterization}
\end{footnotesize}

\noindent Figure \ref{Fig2}e shows an optical microscopy image of the sensor, with the optical field distribution shown in the inset. The microring resonator exhibits an optical $Q$ factor of $1.33 \times 10^6$ in water (Fig. \ref{Fig2}f), which is similar to that of $1.35 \times 10^6$ in air (Extended Data Fig. \ref{FigED1}e), confirming that the packaging robustly preserves the microring optical performance in various operational environments.

The experimental set-up for sensitivity characterization is depicted in Extended Data Fig. \ref{Fig6}. Specifically, a 1550~nm laser is coupled into the microring resonator and detuned to the optimal slope of the optical resonance. A pre-calibrated ultrasound transducer, driven by a vector network analyzer (VNA) and positioned above the sensor, serves as the ultrasound source. The ultrasound-induced optical resonance shift modulates the transmitted light intensity and is detected by a photodetector. The transmission spectrum of the optical mode, noise power spectral density (PSD), and the ultrasound response of the sensor are measured by an oscilloscope, electronic spectrum analyzer (ESA) and VNA, respectively.

Noise PSD across the first-order flapping mode $\nu_{(0,0)}$ is measured in both air and water, with the measured results and Lorentzian fittings shown in grey dots and black curves in Figs. \ref{Fig3}a and \ref{Fig3}b, respectively. Notably, $\nu_{(0,0)}$ mode (with its mechanical profile shown in the insets of Figs. \ref{Fig3}a and \ref{Fig3}b) exhibits the greatest spatial overlap with incident ultrasound from above the sensor and therefore is expected to have better sensitivity. Near the mechanical resonance frequency, thermal noise--arising from both surrounding molecule collisions and intrinsic damping of the mechanical resonator \cite{WGM15_disk}--dominates as Lorentzian peaks (orange dashed curves). Resonance frequencies of the $\nu_{(0,0)}$ mode in air and water are 289~kHz and 52~kHz with linewidths of 10~kHz and 2~kHz, respectively. The resonance frequency shifts downward in water compared to air due to the increased effective mass and viscosity \cite{11}. Away from the mechanical resonance frequency, thermorefractive noise (TRN) dominates (green dashed curves), originating from temperature fluctuations induced refractive index variation via the thermo-optic effect \cite{TRN}. As investigated in previous works, achieving thermal noise-limited performance is essential for enhanced sensitivity \cite{WGM15_disk} (Supplementary Information IA).

The sensor's single-frequency linear dynamic range (LDR) at mechanical resonance is approximately 52~dB in both air and water (Supplementary Information ID), and all subsequent ultrasound pressures are maintained within this linear range. The ultrasound response of the sensor is measured in air and water by sweeping the applied frequency to the transducer using the VNA, with the results shown in red and blue dots in Figs. \ref{Fig3}c and \ref{Fig3}d, respectively. The observed spectral peak matches the mechanical mode identified in Figs. \ref{Fig3}a and \ref{Fig3}b, confirming that the mechanical mode amplifies the membrane's displacement response. As illustrated in the inset of Fig. \ref{Fig3}c, the on-resonance case achieves a signal-noise-ratio (SNR) of 42~dB at 30~Hz resolution bandwidth (RBW), a 7~dB improvement over the off-resonance case (35~dB). Deviations in the off-resonance frequency regions from the fitting curves (red and blue curves in Figs. \ref{Fig3}c and \ref{Fig3}d, respectively) using mechanical susceptibility functions result from interference between responses of two mechanical modes (Supplementary Information section IB). Additionally, the observed response undulations are caused by sound waves reflections between the ultrasound transducer and the sensor chip (Supplementary Information section IIB).

The NEPs of the sensor at different frequencies are derived from the noise spectra, single-frequency response, and response spectra, as shown in Figs. \ref{Fig3}e (air) and \ref{Fig3}f (water). The NEP spectra exhibit minima at mechanical resonance frequencies, consistent with theoretical predictions (Supplementary Information sections IA and IC). Remarkably, our sensor achieves optimal NEPs of 218~nPa/$\sqrt{\rm{Hz}}$ at 289~kHz in air and 9.6~nPa/$\sqrt{\rm{Hz}}$ at 52~kHz in water, representing a record in microcavity-based ultrasonic sensors. The superior sensitivity in water compared to that in air likely arises from improved acoustic impedance match at the water-sensor interface. We also measured the NEPs in a larger frequency range from 20~kHz to 1~MHz, displayed in the insets of Figs. \ref{Fig3}e and \ref{Fig3}f, where the shaded areas correspond to the $\nu_{(0,0)}$ mode, and the pentagrams mark the best sensitivities. Device reproducibility is confirmed across nine sensors from the same batch, showing 1\% variation in mechanical resonance frequencies and 38\% deviations in sensitivities (Supplementary Information IIA). Furthermore, the $\nu_{(0,0)}$ mode demonstrates excellent directional uniformity, with a 3~dB angular bandwidth exceeding 100° (Supplementary Information IIB). Crucially, suspended-membrane sensors outperform non-suspended (without DRIE process) devices by two orders of magnitude in NEP (Extended data Fig. \ref{FigED3}), highlighting the critical role of membrane suspension.

\begin{footnotesize}
\noindent{\bf Photoacoustic gas spectroscopy}
\end{footnotesize}

\noindent Highly sensitive air-coupled ultrasonic sensors are indispensable for gas detection in applications ranging from breath analysis and environmental monitoring to hazardous gas detection \cite{gas1}. Among spectroscopic techniques, photoacoustic spectroscopy (PAS) has emerged as a powerful approach \cite{gas2, gas3}, leveraging the photoacoustic effect where gas molecules convert intensity-modulated light into acoustic waves through periodic thermal expansion. PAS offers unique advantages, including background-free detection, high sensitivity, and broad spectral coverage, making it particularly suitable for trace gas analysis.

We demonstrate the application of our highly sensitive integrated ultrasonic sensor in PAS of C$_2$H$_2$ gas, with the experimental setup shown in Fig. \ref{Fig4}a (details in Extended Data Fig. \ref{FigED4} and Methods). Specifically, a pump laser with its intensity modulated by an electro-optic modulator (EOM), is injected into the gas cell containing a mixture of 1\% C$_2$H$_2$ and 99\% N$_2$, while the integrated ultrasonic sensor within the gas cell detects the generated acoustic waves. PA signals are measured at two modulation frequencies: 288~kHz (on mechanical resonance) and 268~kHz (off resonance), at different pump powers. As the PA signal scales linearly with pump power \cite{gas11}, the noise-equivalent concentration (NEC) is inversely proportional to the pump power (Fig. \ref{Fig4}b). At 350~mW pump power and 1~s integration time, NECs of 2.9~ppm (on resonance) and 13.1~ppm (off resonance) are achieved, demonstrating significant sensitivity enhancement through mechanical resonance. The inset of Fig. \ref{Fig4}b quantifies this improvement, showing SNRs of 61~dB at 288~kHz (on resonance) versus 48~dB at 268~kHz (off resonance). The background-free nature of PAS is confirmed by comparing PA signals from C$_2$H$_2$/N$_2$ mixture and pure N$_2$ (Supplementary Information IIIA), with PA responses exclusively emerging at C$_2$H$_2$ absorption wavelengths. The PA spectrum of our sensor is obtained by scanning the pump laser wavelength with a step of 0.01~nm, as shown in the blue curve of Fig. \ref{Fig4}c, which matches well with the high-resolution transmission molecular absorption database (HITRAN) simulations (red curve). The relative residuals between the two are below 5\% (black curve), validating the system's spectroscopic accuracy.

\begin{footnotesize}
\noindent{\bf Underwater ultrasound imaging}
\end{footnotesize}

\noindent Ultrasound imaging technology is a powerful non-destructive technique widely used in medical diagnostics, industrial inspection, and marine acoustics. We demonstrate these capabilities through underwater ultrasound imaging using our highly sensitive sensors. As depicted in Fig. \ref{Fig4}d, the experimental setup includes a water tank containing a 500~kHz focused ultrasound transducer, our sensor, and an acrylic sample featuring an "F" shaped groove. When immersed in water, the groove traps air, creating a localized acoustic impedance mismatch. The sample is raster-scanned using a two-dimensional horizontal translational stage. When the ultrasound beam generated by the ultrasound transducer is focused on the "F" region, the air-filled groove reflects the ultrasound waves, resulting in significant attenuation of the transmitted ultrasound signal. This generates a strong contrast between the groove and the surrounding homogeneous regions, enabling high-resolution imaging. For optimal performance, we operate the sensor at the $\nu_{(3,0)}$ mechanical mode of 517~kHz (mode profile shown in the inset of Extended Data Fig. \ref{FigED4}b), achieving higher spatial resolution than the $\nu_{(1,0)}$ due to its shorter wavelength.

The sample is scanned with a 1~mm step size, and the sensor’s response is recorded at each position. Figures \ref{Fig4}e(i) and \ref{Fig4}e(ii) represent the imaging results obtained with a driving ultrasound pressure of 0.3~mPa and driving frequencies of 517~kHz (on-resonance) and 550~kHz (off-resonance), respectively, revealing significantly improved contrast and clarity under resonant conditions, a derict consequence of mechanical resonance-enhanced sensitivity. Further quantitative analysis (Supplementary Information IIIB) confirms a spatial resolution of 1.89~mm, demonstrating the potential of our sensor for high-resolution imaging and long-range target detection in underwater applications. Remarkably, our sensor outperforms a commercial hydrophone even at ultrasound pressures three orders of magnitude lower (0.3~mPa versus 0.7~Pa), as shown in Fig. \ref{Fig4}e(iii).

\begin{figure}
\centering\includegraphics{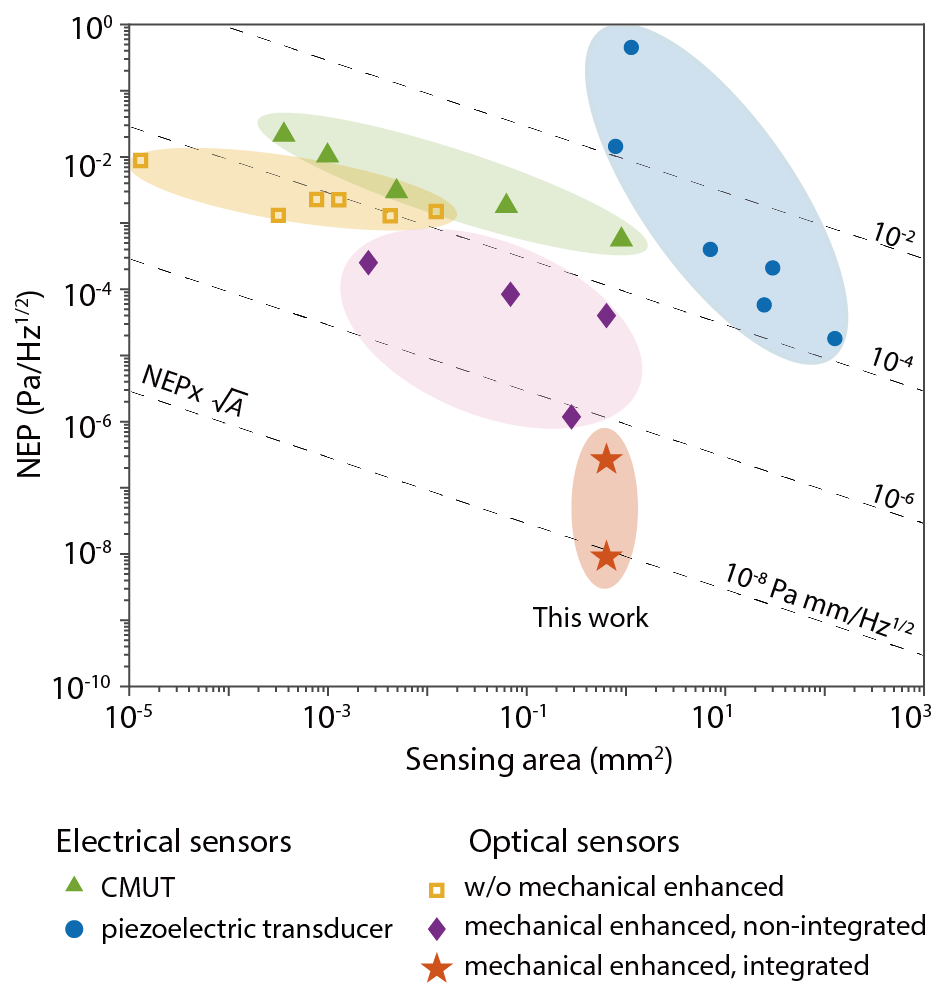}
\caption{\textbf{Performance benchmarking of various ultrasonic sensors.} Noise-equivalent pressures (NEP) as a function of the sensing area $A$. See Supplementary Information IV for data references.} 
\centering
\label{Fig5}
\end{figure}

\medskip
{\bf\noindent Conclusion and discussion}

\noindent Figure \ref{Fig5} compares the NEP versus sensing area $A$ for various ultrasonic sensors. Our sensor achieves the highest sensitivity (lowest NEP) while maintaining a compact footprint. As the NEP of optomechanical sensors scales approximately as $1/\sqrt{A}$, we evaluate the figure of merit NEP$\times \sqrt{A}$ across different technologies. Remarkably, our sensor achieves NEP$\times \sqrt{A}$ values at the 10$^\mathrm{-8}$~Pa mm/Hz$^\mathrm{1/2}$ level, surpassing conventional electric and optical sensors by several orders of magnitude, highlighting its compelling advantages in both miniaturization and sensitivity. A comprehensive performance comparison with existing ultrasonic sensors is provided in Supplementary Information IV.

Several key directions will drive further development of integrated ultrasound sensing technology. First, the sensitivity of the sensor can be further enhanced by geometric optimization. The working bandwidth can be significantly expanded by suppressing photorefractive noise to enable thermal-noise-limited sensitivity across wider frequency ranges and utilizing multiple mechanical modes for broader spectral coverage. Additionally, advances in photonic integrated circuits \cite{chip7, chip8, chip9} enables co-integration of complementary on-chip components, inluding lasers \cite{chip4, chip5}, spectrometers \cite{chip6}, and photodetectors \cite{chip1}, facilitating fully integrated sensor systems. Beyond device optimization, we envision practical applications of these highly sensitive sensors in multimodal sensing and wearable technologies, transitioning these sensitive platforms from laboratory prototypes to field-deployable solutions for mobile sensing \cite{chip2}. Our work establishes a foundation for next-generation compact, high-performance sensing systems with transformative potential across diverse domains, from biomedical diagnostics and industrial monitoring to real-time environmental surveillance.

\bibliography{Main/References}

\bigskip

\noindent\textbf{Methods}

\begin{footnotesize}


\noindent{\bf Device design}

We systematically optimize the device geometry to enhance ultrasound sensitivity (Extended Data Fig. \ref{FigED1}a). The key design parameter is the radius ratio $R_{\rm{ring}}/R_{\rm{mem}}$, where $R_\mathrm{ring}$ and $R_\mathrm{mem}$ denote the radii of the microring and membrane, respectively. Our optimization focuses on the first-order flapping mode $\nu_{(0,0)}$, as it has the largest spatial overlap with the incident ultrasound from above the sensor. The mechanical mode profile of $\nu_{(0,0)}$ is derived by modeling the SiO$_2$ membrane as a thin plate \cite{8}, yielding the normalized displacement:
\begin{equation}
w_{0,0}=J_{0}(\lambda _{00}\frac{R_{\rm{ring}}}{R_{\rm{mem}}})-\frac{J_{0}(\lambda _{00})}{I_{0}(\lambda _{00})}I_{0}(\lambda _{00}\frac{R_{\rm{ring}}}{R_{\rm{mem}}}),
\end{equation}
where $J_{0}$ and $I_{0}$ represent the 0-th Bessel function and the modified Bessel function of the first kind, respectively. $\lambda _{00}$ is a dimensionless parameter which is related to the mode shape, determined from Ref. \cite{14}. The normalized radial change $\Delta R$ is calculated as \cite{9}:
\begin{equation}
\Delta R=-H_\mathrm{pos}\frac{\partial w_{0,0}}{\partial R_{\rm{ring}}},
\end{equation}
where $H_\mathrm{pos}$ is the vertical position of the waveguide. Extended Data Fig. \ref{FigED1}b shows $\Delta R$ versus $R_{\rm{ring}}/R_{\rm{mem}}$, revealing a maximum at $R_{\rm{ring}}/R_{\rm{mem}}$=0.52. Considering an EBL write field of 500~\textmu m, we implement $R_{\rm{ring}}$=450~\textmu m and $R_{\rm{mem}}$=235~\textmu m.

We systematically investigate the intrinsic $Q$ factor of a Si$_3$N$_4$ microring resonator with a radius of 235~\textmu m and a thickness of 265~nm, as a function of microring width and SiO$_2$ cladding thickness (Extended Data Figs. \ref{FigED1}c,d). The intrinsic $Q$ factor exhibits strong geometric dependence, showing a sharp decline for microring width below 6~\textmu m (Extended Data Figs. \ref{FigED1}c) while remaining high for cladding thickness above 2~\textmu m (Extended Data Figs. \ref{FigED1}d). While thicker cladding enhances optical $Q$ factor, it comprises mechanical compliance and thus decreases sensitivity. We therefore optimize the design with 6~\textmu m microring width and 2~\textmu m cladding thickness, achieving both intrinsic $Q$ factors (1.35$\times 10^6$ in air, Fig. \ref{FigED1}e) and environmental stability without sacrificing sensitivity.

\medskip

\noindent{\bf Device fabrication}

The ultrasonic sensors are fabricated on a 4-inch wafer comprising a 265~nm-thick Si$_3$N$_4$ layer on a 4~\textmu m-thick buried-oxide layer above a 500~\textmu m-thick silicon substrate. The fabrication begins with patterning Si$_3$N$_4$ microring resonators and bus waveguides using electron beam lithography (EBL) with AR-P 6200 resist, followed by reactive ion etching (RIE) employing CHF$_3$/O$_2$ plasma. A protective 2-\textmu m-thick SiO$_2$ cladding layer is subsequently deposited via inductively coupled plasma chemical vapor deposition (ICP-CVD) to shield the microring resonator from environmental contamination. We then flipped the chip upside down and patterned the SiO$_2$ membrane by double-sided aligned photolithography 
To ensure precise concentric alignment between the Si$_3$N$_4$ microring and SiO$_2$ membrane structures, we perform double-sided aligned photolithography after flipping the wafer upside down. The SiO$_2$ membrane release is achieved through a Bosch process utilizing alternating C$_4$F$_8$ passivation and SF$_6$ etching cycles to remove the underlying silicon substrate. Finally, the completed devices are diced and packaged with mode conversion fibers to create compact, portable sensor units.

\medskip

\noindent{\bf Ultrasound sensitivity characterization}

A 1550~nm laser (TOPTICA CTL1500) is coupled into the microring resonator, with its frequency detuned relative to the optical mode to maintain 25\% transmission power for optimal sensitivity \cite{3}. The transmitted optical signal is detected using a photodetector (KEYANG KY-PRM-10M-1-FC) and recorded via an oscilloscope. Ultrasound waves are generated by a pre-calibrated ultrasound transducer \cite{WGM29_disk}, positioned above the sensor. Ultrasound pressure calibration at the sensor location is performed using: (i) a needle hydrophone (Onda HNR-1000) for aqueous measurements, and (ii) a scanning laser vibrometer (Sunnyinnovation optical Intelligence LV-SC400) for measurements in air. For single-frequency response characterization, the ultrasound transducer is driven by a sinusoidal signal generated by an arbitrary function generator, with the sensor response recorded using an electrical spectrum analyzer (ROHDE\&SCHWARZ FPS4). Broadband frequency response of the sensor is measured via a vector network analyzer (ROHDE\&SCHWARZ ZNL3), which sweeps ultrasound frequencies while recording the corresponding sensor response.

\medskip

\noindent{\bf Experimental details for photoacoustic gas spectroscopy}

Photoacoustic gas spectroscopy measurements are performed at room temperature and atmospheric pressure using a sealed gas cell with fiber-optic feedthroughs for optical access. 
The cell features a gas inlet port connected to acetylene (C$_2$H$_2$) and nitrogen (N$_2$) cylinders for precise gas mixing, and an outlet port connected with a built-in barometer for pressure monitoring. Before measurements, the gas cell is purged with 1\% C$_2$H$_2$ mixed with 99\% N$_2$, by opening the C$_2$H$_2$ supply valve (valve 1) and vent valve (valve 2) to ensure complete air displacement and concentration accuracy. During measurements, both valves are closed to maintain a static gas environment, with identical procedures followed for N$_2$ measurements. A CW pump laser (TOPTICA CTL1550) is used to excite the PA signals of the gas, with its intensity modulated by an electro-optic modulator (Exail MX-LN-10. The modulation frequency is controlled by a function generator, while the modulation depth is adjusted via a DC voltage. Before entering the gas cell, the pump power is amplified by an erbium-doped fiber amplifier (Amonics AEDFA-37-R-FA) and converted to a free-space beam with a 1-2~mm diameter spot using a fiber collimator. The integrated ultrasonic sensor, positioned below the optical path, detects the resulting PA signal of the gas.

\end{footnotesize}

\medskip

\noindent\textbf{Data availability}

\begin{footnotesize}
\noindent The data that support the plot within this paper and other findings of this study are available upon publication. 
\end{footnotesize}

\medskip

\noindent\textbf{Code availability}

\begin{footnotesize}
\noindent The codes that support the findings of this study are available upon publication. 
\end{footnotesize}

\medskip 

\noindent\textbf{Acknowledgments}
\begin{footnotesize}
\noindent The authors thank Professors Yun-Feng Xiao, Zhixiong Gong, and Xue-Feng Jiang for useful discussions, and Prof. Tuo Liu for calibrating the ultrasound pressure in air. This work was supported by the Synergetic Extreme Condition User Facility (SECUF, https://cstr.cn/31123.02.SECUF). 

Funding: Innovation Program for Quantum Science and Technology (2023ZD0301100); National Natural Science Foundation of China (NSFC) (62222515, 12174438, 11934019, 12274438); The National Key Research and Development Program of China (2021YFA1400700, 2021YFB3501400); Basic frontier science research program of Chinese Academy of Sciences (ZDBS-LY-JSC003); CAS Project for Young Scientists in Basic Research (YSBR-100).
\end{footnotesize}
\medskip

\noindent\textbf{Author contributions} 
\begin{footnotesize}
\noindent Experiments were conceived and designed by X.C., H.Y., M.W., B-B.L., and Q-F.Y. Measurements and data analysis were performed by X.C., H.Y., and M.W., with assistance from Z-G.H. and Z-L. W. Devices were designed by X.C., H.Y., and M.W. Devices were fabricated by X.C., H.Y., and M.W., with assistance from Y.W. and C.L. The project was supervised by B-B.L. and Q-F.Y. All authors participated in preparing the manuscript.
\end{footnotesize}

\medskip

\noindent\textbf{Competing interests}

\begin{footnotesize}
\noindent The authors declare no competing interests.
\end{footnotesize}

\medskip

\noindent\textbf{Additional information}

\renewcommand{\figurename}{\textbf{Extended Data Fig.}}
\begin{footnotesize}
\noindent \textbf{Supplementary information} is available for this paper.

\medskip

\noindent \textbf{Extended data} is available for this paper.

\medskip

\noindent {\bf Correspondence and requests for materials}  should be addressed to B-B.L.
\end{footnotesize}

\setcounter{figure}{0}

\begin{figure*}
\centering\includegraphics[width=160mm]{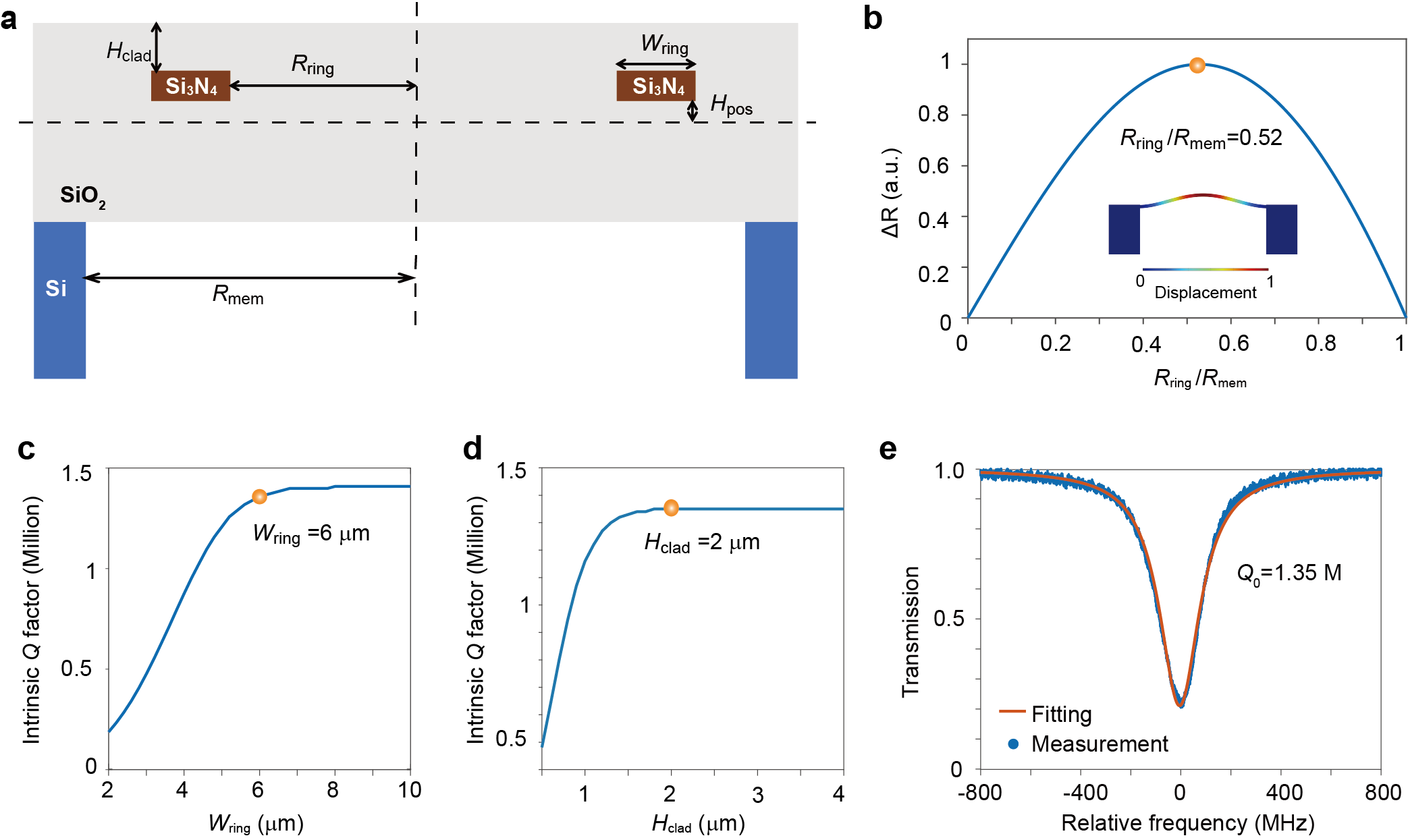}
\caption{\textbf{Device design.}
\textbf{a} Schematic of the sensor with key geometric parameters.
\textbf{b} Microring radius variation $\delta R$ for different microring-to-membrane radius ratios $R_\mathrm{ring}/R_\mathrm{mem}$. Inset: Cross-sectional profile of the membrane's first-order flapping mode $\nu_{00}$.
\textbf{c,d} Simulated intrinsic $Q$ factor of the Si$_3$N$_4$ microring resonator in water, as functions of (\textbf{c}) the microring width and (\textbf{d}) Si$_2$silica cladding thickness.
\textbf{e} Measured transmission spectrum of an optical resonance of the Si$_3$N$_4$ microring resonator (blue dot), along with its Lorentzian-fitting (red curve), indicating an intrinsic optical $Q$ factor of 1.35 $\times$ 10$^6$ in air. }
\label{FigED1}
\end{figure*}

\begin{figure*}[ht!]
\centering\includegraphics{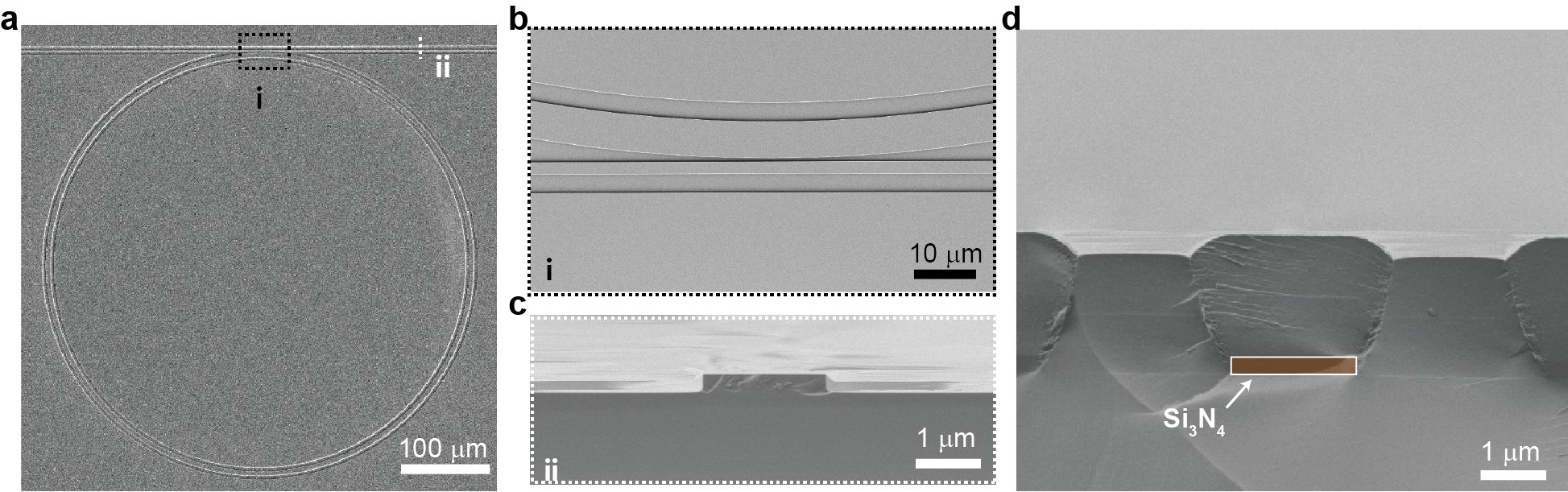}
\caption{\textbf{Sensor characterization.}
\textbf{a} Scanning electron micrograph (SEM) of the Si$_3$N$_4$ microring resonator before SiO$_2$ cladding layer deposition. 
\textbf{b} SEM of the etched Si$_3$N$_4$ microring and its coupling section with the adjacent bus waveguide.
\textbf{c} Cross-sectional view SEM of Si$_3$N$_4$ waveguide structure atop the SiO$_2$ undercladding layer.
\textbf{d} SEM of Si$_3$N$_4$ waveguide core fully encapsulated by the SiO$_2$ cladding layer. The false-colored region highlights the Si$_3$N$_4$ waveguide.
 }
\label{FigED2}
\end{figure*}

\begin{figure*}
\centering\includegraphics[width=70mm]{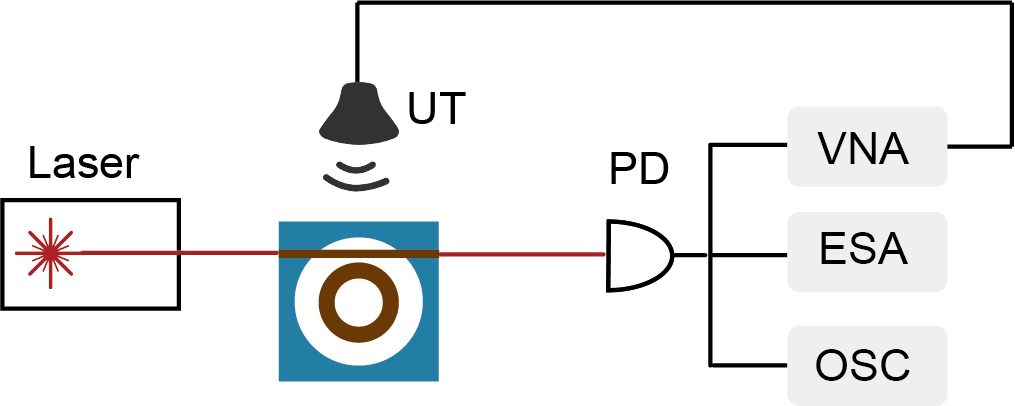}
\caption{\textbf{Experimental setup for ultrasound sensitivity measurements.} 
UT: ultrasound transducer; PD: photodetector; OSC: oscilloscope; ESA: electronic spectrum analyzer; VNA: vector network analyzer.}
\label{Fig6}
\end{figure*}

\begin{figure*}
\centering\includegraphics[width=80mm]{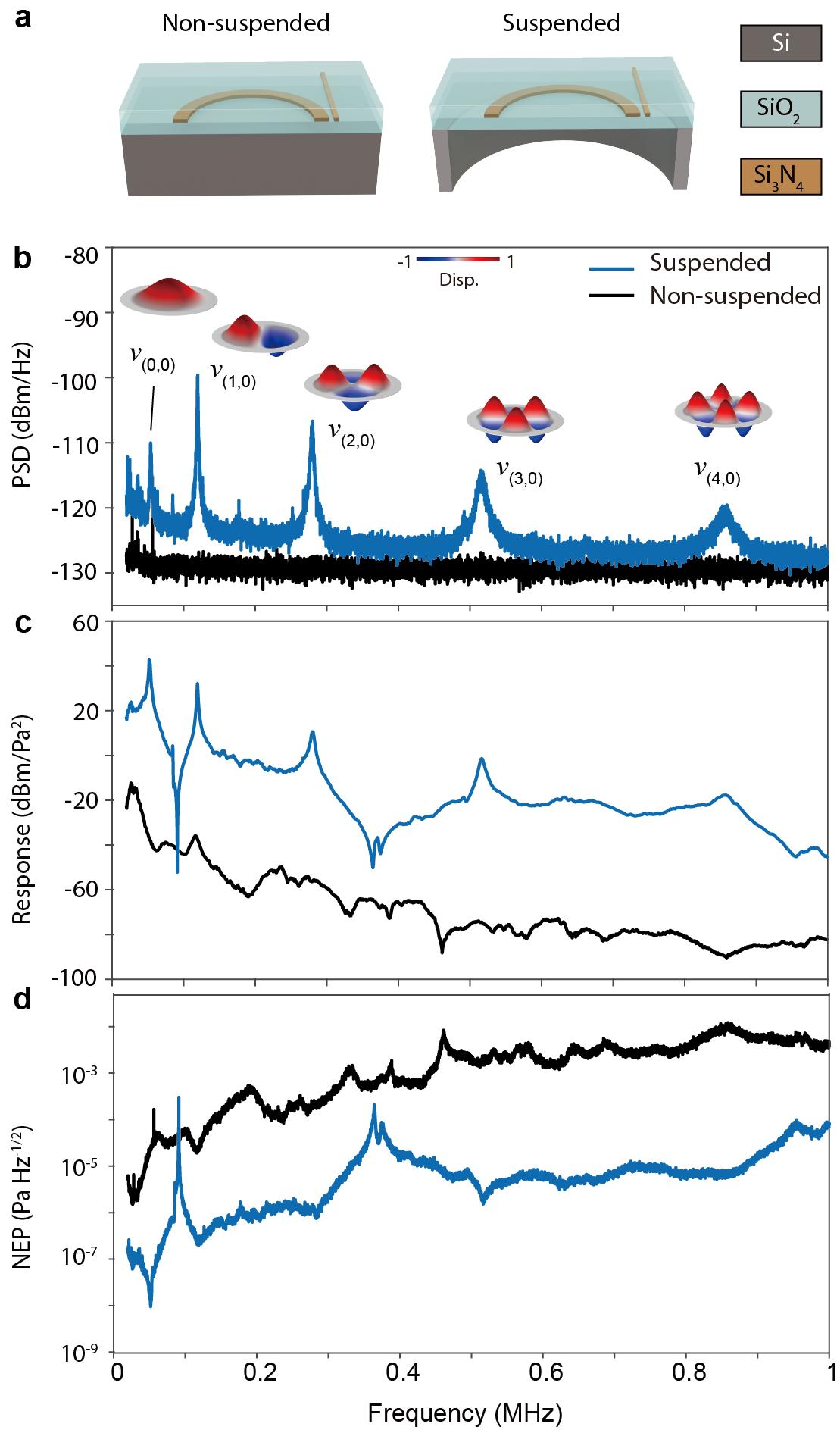}
\caption{\textbf{Performance comparison of non-suspended and suspended sensor structures in water.}
\textbf{a} Schematic illustrations of the non-suspended (left) and suspended (right) sensor configurations.
\textbf{b-d} Comparative results of noise power spectra densities (PSDs) (\textbf{b}), frequency-dependent ultrasound responses (\textbf{c}), and noise-equivalent pressures (NEPs) (\textbf{d}) of the sensors with both non-suspended (black curves) and suspended (blue curves) structures.}
\label{FigED3}
\end{figure*}

\clearpage

\begin{figure*}
\centering\includegraphics[width=160mm]{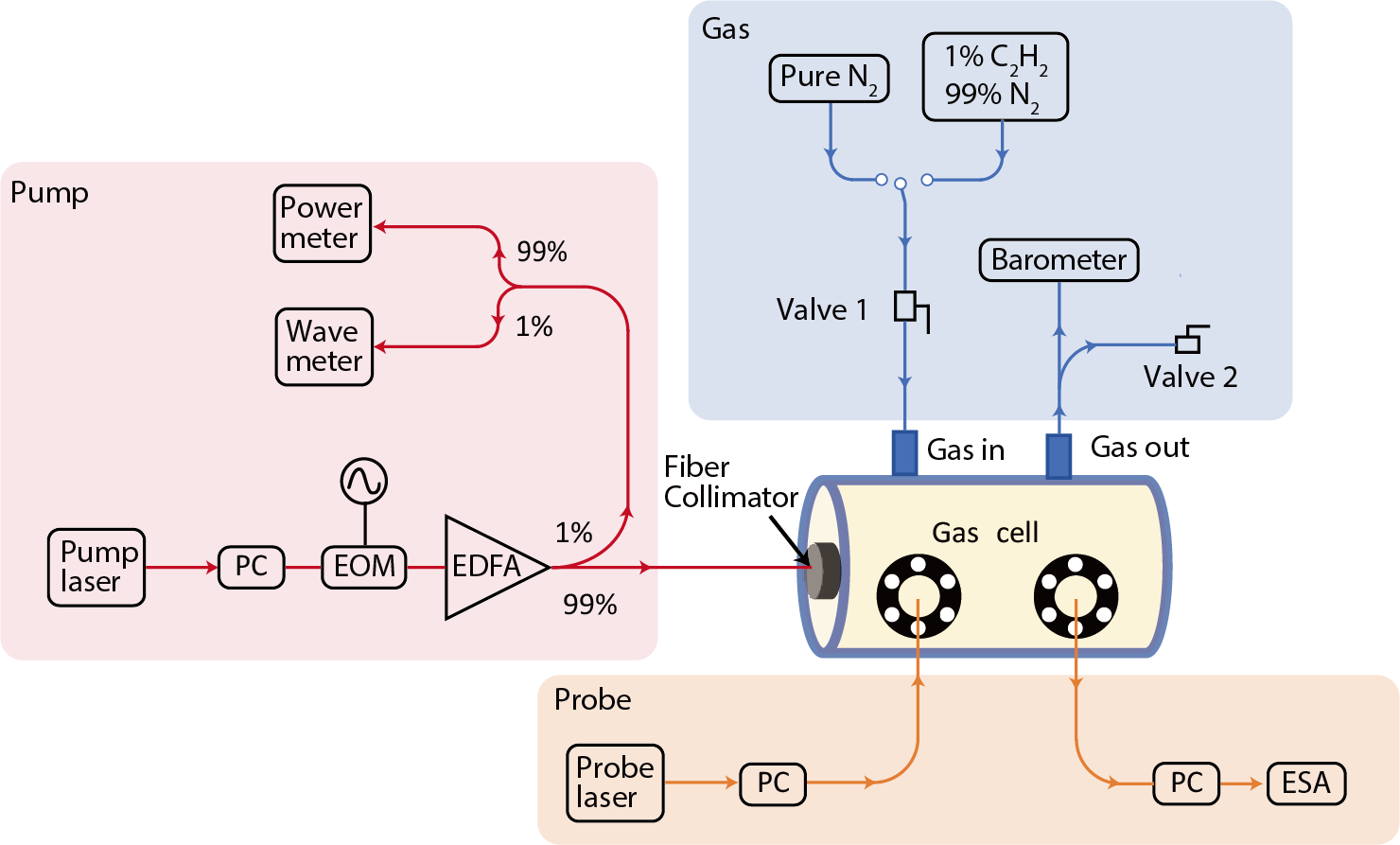}
\caption{\textbf{Experimental setup for the photoacoustic gas spectroscopy measurement.} PC: polarization controller; AFG: arbitrary function generator; EOM: electro-optical modulator; EDFA: erbium-doped fiber amplifier; PD: photodetector; ESA: electronic spectrum analyzer.}
\label{FigED4}
\end{figure*}

\end{document}